\documentclass[epj]{svjour}

\usepackage{graphics}
\usepackage{tikz}
\usetikzlibrary{calc}
\usepackage{psfrag}
\usepackage{amsmath,amsfonts,amssymb}

\begin{document}
\title{Turbulent spots in channel flow: an experimental study}
\subtitle{Large-scale flow, inner structure and low order model}
\author{
	Gr\'egoire Lemoult\inst{1} \and
	Konrad Gumowski\inst{2} \and
	Jean-Luc Aider\inst{1} \and
	Jos\'e Eduardo Wesfreid\inst{1}
}

%
%
\institute{
Laboratoire de Physique et M\'ecanique des Milieux H\'et\'erog\`enes (PMMH), UMR CNRS 7636, ESPCI, UPMC, Paris Diderot, 10 rue Vauquelin, 75005 Paris, France \and
Warsaw University of Technology, Inst Aeronaut \& Appl Mech, PL-00665 Warsaw, Poland 
}

\date{Received: date / Revised version: date}
%
\abstract{
We present new experimental results on the development of turbulent spots in channel flow. The internal structure of a turbulent spot is measured, with Time Resolved Stereoscopic Particle Image Velocimetry. We report the observation of travelling-wave-like structures at the trailing edge of the turbulent spot. Special attention is paid to the large-scale flow surrounding the spot. We show that this large-scale flow is an asymmetric quadrupole centred on the spot. We measure the time evolution of the turbulent fluctuations and the mean flow distortions and compare these with the predictions of a nonlinear reduced order model predicting the main features of subcritical transition to turbulence.
\PACS{
      {PACS-key}{discribing text of that key}   \and
      {PACS-key}{discribing text of that key}
     } 
} 
%


\maketitle

\section{Introduction}
\label{intro}

Transition to turbulence in wall bounded shear flows, like plane Couette (PCF), circular Poiseuille (CPF), plane \\ Poiseuille (PPF) or boundary layer flow, is one of the most intriguing examples of complex and disordered behaviour in nature. The study of hydrodynamic instabilities and dynamical systems theory involve in this transition is an important field of research \cite{manneville1995dissipative}.
Turbulence appears abruptly, not through a sequence of transitions, and is localized into turbulent spots, surrounded by laminar flow. These spots, first described by Emmons \cite{emmons1951laminar}, are not composed of a single large-scale homogeneous structure, but of an assemblage of small-scale vortices, separated from laminar flow by fronts.

The dynamics of subcritical transition to turbulence is strongly related to the existence and interplay of these localized spots. Other isolated structures have been observed in a great variety of physical systems \cite{dawes2010emergence}, such as thermal convection \cite{knobloch2008}, magnetic liquids \cite{richter2005two}, granular material \cite{umbanhowar1996localized} or buckling instabilities \cite{champneys1999localization}. Indeed, localized structures are related to subcritical instabilities, \textit{i.e.} due to finite amplitude perturbations.

We are interested in the case of transition to turbulence of plane Poiseuille flow in a channel of rectangular cross section. Channel flow can be described by a single dimensionless parameter, the Reynolds number $Re=U_{cl}h/\nu$, where $U_{cl}$ is the center line velocity, $h$ the channel half height and $\nu$ the kinematic viscosity of the fluid. 
Linear stability theory predicts undisturbed stable laminar Poiseuille flow until $Re = 5772$ \cite{orszag1971accurate}, but experiments show transition at Reynolds numbers near $Re \approx 1300$ \cite{carlson1982,klingmann1992,lemoult2012}, in the form of  isolated turbulent spots. 
Research on turbulent spots in channel flow has been performed with flow visualisation \cite{carlson1982,alavyoon1986turbulent}, local measurements \cite{klingmann1992,seki2012experimental} and numerical simulations in both PPF \cite{henningson1987numerical,henningson1991turbulent,tsukahara2005dns,tsukahara2011,kawahara2012,tuckerman2013ETC14} and PCF \cite{lundbladh1991,schumacher2001evolution,lagha2007modeling,duguet2013oblique}.

In the present study, we present new experiments which provide quantitative measurements of the full velocity field of a turbulent spot,  as has been performed in pipe flow by van Doorne and Westerweel \cite{vandoorne2009flow} and recently by Lemoult \textit{et al.} \cite{lemoult2013turbulent} in channels.
The fine structure of the velocity field is measured by means of Time Resolved Stereoscopic Particle Image Velocimetry (TR-SPIV). A Taylor's hypothesis is assumed to reconstruct the three dimensional field from a time series.

Turbulent spots shows transient growth, after which they either decay or are sustained, with a complex spatio-temporal intermittent behaviour. Many questions remain open about the growth (streamwise expansion), the spreading (spanwise expansion), the splitting and the interaction of these turbulent domains, surrounded by laminar ones. 
The inhomogeneity of flow friction generates a coupling between the flow at the intermediate scale inside the turbulent domain and an external induced large-scale flow which can influence the morphology of turbulent spots. This topic has been studied theoretically and numerically, especially in PCF \cite{schumacher2001evolution,lagha2007modeling}, but not yet observed experimentally until very recently in PPF \cite{lemoult2013turbulent}.
The spot spreading can be seen as a consequence of this large-scale flow as pointed out by Duguet \textit{et al.} \cite{duguet2013oblique} in PCF in relation with the random nucleation of new streaks in the vicinity of the turbulent-laminar boundary \cite{duguet2011stochastic}.
It has been observed in PPF that the leading and trailing edges of spots expand at different velocities \cite{carlson1982,alavyoon1986turbulent} but the mechanism behind these observations remains unclear. Shimizu and Kida \cite{shimizu2009driving}, Duguet \textit{et al.} \cite{duguet2010slug} and Hof \textit{et al.} \cite{hof2010eliminating} highlighted the local instability which occurs at the trailing edge as a driving mechanism of the spot. In certain cases, in PPF and PCF, turbulent spots show a complex spatio-temporal behaviour, including collective organisation and branching in the form of bands \cite{prigent2003long,barkley2005computational,manneville2011decay}.

In order to understand the existence of turbulent spots, many theoretical efforts had been developed. The non-normal linear stability theory explains that inner structures present maximum transient growth when they are oriented in the direction of the flow as streamwise streaks \cite{trefethen1993,reddy1993energy}. Linear stability theory also predicts the breakdown of these streaks \cite{reddy1998}. However linear models fail to predict the existence of a sustained state in which laminar and turbulent areas coexist. A preferred framework for understanding this situation is provided by nonlinear models of self-sustained turbulence \cite{waleffe1997,shimizu2009driving,hall2010streamwise}.
If sufficiently strong streamwise streaks exist inside the turbulent spot, a longitudinal wavy instability occurs. This streamwise modulation of the streaks will regenerate initial streamwise rolls through a non linear mechanism. If this is the case, nonlinear amplification occurs, in time and space, inside the  spot leading to a turbulent state.

According to dynamical systems theory the disordered dynamics of turbulence as well as of its edge are organized around unstable solutions of the Navier-Stokes equations. Within the past two decades, the computation of exact solutions of the Navier-Stokes equation has attracted considerable attention. The discovery of these exact coherent structures has opened a new approach to understanding the dynamics of unsteady flows in transitional $Re$ range. Starting with the computation by Nagata \cite{nagata1990three} of the first unstable three dimensional nonlinear equilibrium solution of PCF, a large number of equilibria and travelling waves of PCF and CPF have since been found. For a review of the subject, the reader is invited to refer to Kawahara \textit{et al.} \cite{kawahara2012significance}. A few of these were found in PPF \cite{waleffe2001exact,itano2001dynamics,nagata2013preprint,zammert2013}. Due to the unstable nature of these solutions, it is non-trivial to observe them experimentally. Nevertheless recently, de Lozar \textit{et al.} \cite{de2012edge} computed an exact solution of the Navier Stokes equation in CPF using experimental data as an initial guess for an iteration method.

Over the past years, many minimal or phenomenological models have been suggested to capture the main features of the transition to turbulence in wall bounded shear flows, such their subcritical character with finite amplitude perturbations and the existence of localized domains. Some of them are based on quintic amplitude equations, such extensions of Landau-Ginzburg or Swift-Hehenberg equations \cite{pomeau1986front,berge1984ordre,sakaguchi1996stable}. Other nonlinear models, as developed by Waleffe \cite{waleffe1997}, are derived from a strong modal reduction of the Navier-Stokes PDEs into ODEs. These models introduce a separation between a weak spanwise fluctuations (rolls), streamwise fluctuations (streaks), streamwise waviness of streaks and base-flow distortion. Spatial extensions of this model include diffusive terms \cite{dawes2011turbulent,manneville2012turbulent} in order to generate spontaneous patterns.
Recently, Barkley \cite{barkley2011simplifying} proposed a more compact model for transitional CPF with only two variables: the turbulent fluctuations and the base flow distortion. He modelled the transitional pipe flow as a one-dimensional excitable and bistable medium. His model captures the main feature of transition in pipe flow as metastability of localized puffs, puff splitting and slugs.

The purpose of this article is to provide a precise experimental description of the flow field in and around a turbulent spot in PPF during its genesis and to discuss the relevance of low order models in predicting the main features of subcritical transition to turbulence in channel flow.

\section{Experimental set-up}
\label{setup}

\begin{figure}
\begin{center}
\includegraphics[width=\columnwidth]{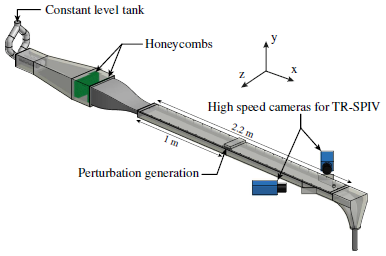}
\caption{\label{fig:setup}\textbf{a.}~Schematic view of the water channel. The development section is 1 m long and the test section is 1.2 m long. We use two synchronized high speed cameras to measure the velocity field in the plane $x=85h$ in a $(\Delta y,\Delta z) = (2h,15h)$ area.}
\end{center}
\end{figure}
 
The experimental system is composed of a 3-m-long plexiglass, constant-pressure-driven channel (figure~\ref{fig:setup}). The test section's half height is $h = 10$~mm, its length is $220h$, and its width is $2L_z = 15h$. The Reynolds number is defined as $Re=U_{cl,lam}h/\nu$, where $U_{cl,lam}$ is the center line velocity measured in absence of the spot. The $x$, $y$ and $z$ axes are, respectively, the streamwise, wall normal and spanwise coordinates, with $y = 0$ in the middle of the channel and $(x,z) = (0,0)$ where perturbations are injected. We define the moving coordinate $x^*=x/h - t.(2/3)U_{cl,lam}/h$, where $(2/3)U_{cl,lam}$ is the mean, or bulk, velocity along the wall normal coordinate on the laminar case. The design of the inlet section, together with the smooth connections between all parts of the channel, minimize the upstream perturbations and keep the flow laminar up to $Re = 5500$, which corresponds to the maximum free-stream velocity of this channel. The perturbation is generated $100h$ downstream from the inlet to ensure a fully developed Poiseuille flow.

The flow is perturbed by a round water jet normal to the flow, with diameter $d = 0.2 h$, drilled into the upper wall. In the following, we will study the response of the flow to a single, short perturbation ($\Delta t = 150$ ms). The structure of the flow induced by the jet may be complex and depends strongly on the amplitude of the perturbation. Nevertheless, above a critical value and above $Re\approx 1300$, this small perturbation will always trigger the development of a turbulent spot \cite{lemoult2012}.

\begin{figure}
	\centering
	\includegraphics[width=\columnwidth]{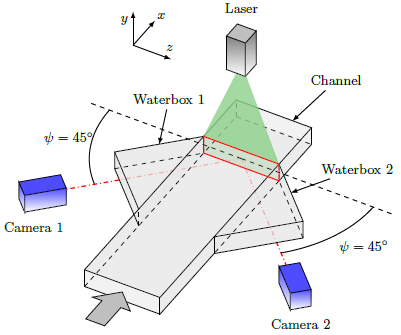}
	\caption{Sketch of the experimental apparatus used for Time Resolved Stereoscopic PIV. Two high speed cameras are positioned on opposite lateral sides of the channel at $45^\circ$ to the observation plane.}
	\label{fig:Stereo}
\end{figure}

A Time-Resolved Stereoscopic Particle Image Velocimetry (TR-SPIV) system is used to measure the three components of the velocity fields $(U,V,W)$ in the $x=85h$ plane (figure~\ref{fig:Stereo}). The fluid is seeded with neutrally buoyant tracer particles ($d_p\approx 20 \mu$m) and a cross-sectional plane is lit by a laser sheet. This plane is viewed by two high speed cameras (Phantom v9) positioned on opposite lateral sides of the channel. The angle between the observation plane and the light sheet plane is set to $45^\circ$. We also use Scheimplfung adapters to ensure the focus of particles in the entire channel cross section. In order to minimize aberrations due to the air/plexiglass dioptre, two water-boxes are added. The passage of turbulent structures is recorded in a series of 1500 contiguous measurements at sampling frequencies between 100 and 200 Hz. The frame rate of the cameras has been adapted for each $Re$ to ensure 10 frames per advective time unit ($h/U_{cl}$). The full three-component velocity field is reconstructed using the commercial software Davis from LaVision. We start the PIV calculation with interrogation windows of $64\times64$ px and decrease the size until $16\times 16$ px with an overlap of 50\%. Two passes are done for each sizes. This calculation give a final spatial resolution of $(N_y \times N_z)=(41 \times 301)$ points in the cross sectional plane. To a first approximation, the spatial structure can be recovered from the temporally resolved measurement by multiplication with the mean advection speed of the flow structures (Taylor's frozen turbulence hypothesis). Because of the fast downstream advection, structures change little while they move over short distances (order of $O(h)$), justifying this hypothesis.

\section{Results and discussion}
\label{results}

\begin{figure}
\centering
\includegraphics[width=\columnwidth]{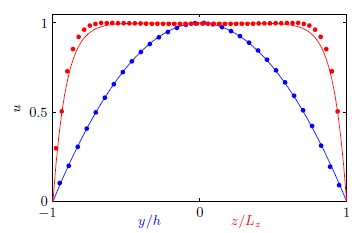}
\caption{\label{fig:Profil_Lam}Laminar velocity profile measured by stereoscopic PIV for $Re=2000$. Solid lines are the theoretical profile and points are experimental data. In blue, profile measured in the $z=0$ plane, $U_{lam}(x=85h,z=0)$, and in red, profile measured in the $y=0$ plane, $U_{lam}(x=85h,y=0)$, non-dimensionalized  by $U_{lam}(y=0,z=0)$.}
\end{figure}

The measured velocity field $\textbf{U} = (U,V,W)$ is decomposed as the sum of its laminar component $\textbf{U}_{lam}$ and a perturbation $\textbf{u}$. 
\begin{equation}
\mathbf{U} = \mathbf{U}_{lam} + \mathbf{u}
\end{equation}
The laminar part of the flow, $\mathbf{U}_{lam}(y,z)$, is obtained by averaging the velocity field before the passage of the spot over 100 measurements.
Figure \ref{fig:Profil_Lam} shows the laminar part of the velocity field, $U_{lam}(y=0)$ and $U_{lam}(z=0)$, measured for $Re=2000$ and adimensionalized by $U_{cl,lam}=U_{lam}(y=0,z=0)$. This measured laminar velocity profile is compared to the theoretical one, plotted as solid lines in figure \ref{fig:Profil_Lam}. The Poiseuille parabolic velocity profile is recovered along the $y$-axis. Along the $z$-axis the profile deviates slightly from the theoretical one, especially close to the walls. This means that lateral boundary layers are not fully developed. This last point which could be seen as negative, is in fact valuable since the parabolic profile is valid in a larger area in the channel. Indeed, the profile along the $y$-axis remains parabolic for $-6h<z<6h$.

\subsection{Spot structure}

\begin{figure*}
\centering
\includegraphics[width=\textwidth]{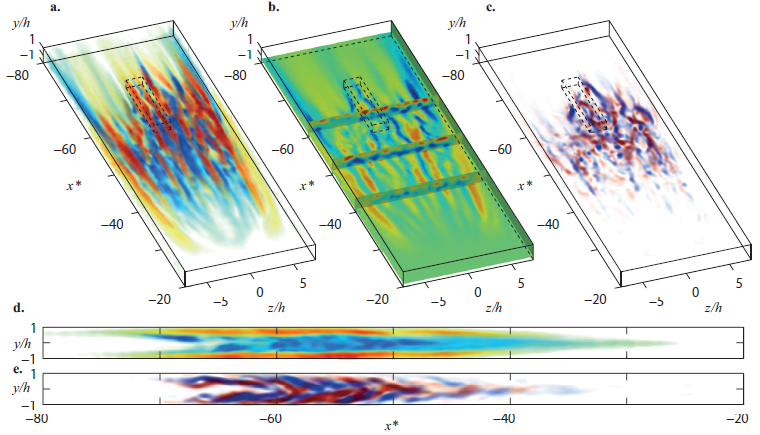}
\caption{\label{fig:Spot}Reconstructed 3D velocity field, assuming Taylor's hypothesis, of a spot at $Re=1500$. \textbf{a.} and \textbf{d.} Streamwise fluctuation $u$ as a volumetric visualization (a color and a transparency is associated with each voxel) from $u=-0.2 U_{cl,lam}$ (blue) to $u=0.2U_{cl,lam}$ (red) and $u=0$ is transparent. \textbf{b.} Slice of streamwise fluctuations in the plane $y=-0.5h$ and slices in the planes $x^*=-40, -50 \mbox{ and } -60$. \textbf{c.} and \textbf{e.} Streamwise vorticity $\omega_x$ as a volumetric visualization from $\omega_x=-\omega_{max}$ (blue) to $\omega_x=\omega_{max}$ (red) and $\omega_x=0$ is transparent. The dotted box represented in \textbf{a.-b.-c.} correspond to the region of space shown in figure \ref{fig:TravellingWaves}.}
\end{figure*}

Figure \ref{fig:Spot} shows a typical reconstruction of the 3D velocity field assuming Taylor's hypothesis. The spot represented in this figure has been recorded at $Re=1500$. Figures \ref{fig:Spot}.a, \ref{fig:Spot}.b and \ref{fig:Spot}.d show the streamwise velocity $u$ and figures \ref{fig:Spot}.c and \ref{fig:Spot}.e represents the streamwise vorticity $\Omega_x$. The flow is dominated by elongated streamwise streaks. Fast velocity streaks are located close to the wall and are well localized, \textit{i.e.} their positions in the $y-z$ plane do not vary much in time. Low velocity streaks on the other side are more mobile and are located in the central part of the channel.

From figures \ref{fig:Spot}.c and \ref{fig:Spot}.e, we observed that streamwise vorticity $\Omega_x$ is concentrated in the upstream half of the spot. Streamwise vortices present in the flow are localised in a $x-y$ plane and are tilted relatively to the streamwise coordinate $x$ with an angle of around $\pm 5^\circ$, estimated from the auto-correlation function. In the downstream part of the spot (leading edge), vortices are concentrated in the centre of the channel whereas close to the trailing edge they form two layers of vortices close to each wall.

\begin{figure}
\centering
\includegraphics[width=\columnwidth]{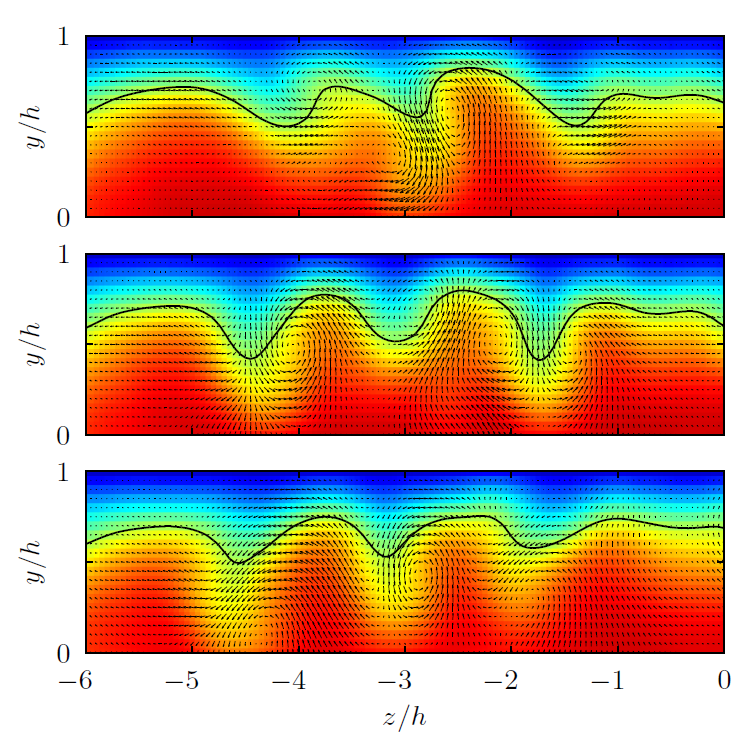}
\caption{\label{fig:Sinuous} Sequence of snapshots at $x*=-57.4, -55.8 \mbox{ and } -54.2$ showing a sinuous instability observed in a spot recorded at $Re=1250$. Color is the streamwise velocity $U$ from 0 (blue) to 1 (red) and arrows represent the in-plane motion $(V,W)$. The solid line shows the $U=0.5$ iso-line. }
\end{figure}

\begin{figure}
\centering
\includegraphics[width=\columnwidth]{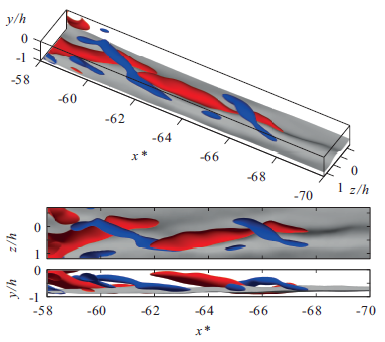}
\caption{\label{fig:TravellingWaves}Travelling wave like structure recorded close to the trailing edge of the spot represented in figure \ref{fig:Spot} (see box). In blue and red, iso-surfaces of streamwise vorticity $\omega_x=-0.25*h/U_{cl,lam}$ (blue) and $\omega_x=+0.25*h/U_{cl,lam}$ (red). In gray, iso-surface of streamwise velocity $U=0.5 U_{cl,lam}$.}
\end{figure}

Although the inner structure of a turbulent spot is highly stochastic, TR-PIV allows us to capture the instantaneous 3D field of velocity inside a spot. This instantaneous picture allows us to extract some very transient features of shear flow turbulence. Figures \ref{fig:Sinuous} and \ref{fig:TravellingWaves} highlight some features which can be found inside a turbulent spot. One key component of the self sustaining process in transitional shear flows is based on the assumption that there exists an instability of the streaks which regenerates streamwise modes. Figure \ref{fig:Sinuous} represents three snapshots of a part of the flow recorded in a $Re=1250$ turbulent spot. This sequence suggests the existence of a sinuous instability which occurs inside the spot.

An important advance in the understanding of the transition to turbulence in past years has been the discovery of travelling wave solutions in shear flows. Figure \ref{fig:TravellingWaves} shows a zoom inside the box, represented in figure \ref{fig:Spot}, close to the trailing edge of this spot. In this figure we represent by blue and red iso-surfaces of streamwise vorticity $\Omega_x$ and in gray the iso-surface of $U=0.5$. From the bottom view, we see clearly the tilt of streamwise vortices in the $y-z$ plane. The middle view give a clear insight into the alternating pattern which streamwise vortices form around the low-speed streak visualized by the bump in the $U$ isosurface in gray. This travelling wave like structure is in a really good agreement with travelling wave solution found in PPF \cite{itano2001dynamics,nagata2013preprint,gibson2013spatially,zammert2013}. Travelling wave like structures are also observed for other $Re$ between 1250 and 2000. Above $Re=2000$, coherent structures subsist but are more difficult to distinguish. To our knowledge, this is the first experimental observation of structures resembling nonlinear travelling waves in PPF.

\subsection{Large-scale flow}
\label{par:LSflow}

\begin{figure}
\centering
\includegraphics[width=\columnwidth]{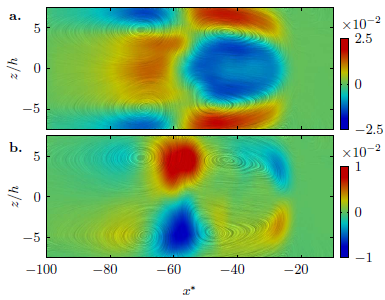}
\caption{\label{fig:LargeScaleMean-xz}Large-scale flow averaged over $-h<y<h$ for $Re=1500$ textured by LIC technique. \textbf{a.} Streamwise component of the large scale flow, $\langle u_{LS} \rangle_y /U_{cl,lam}$. \textbf{b.} Spanwise component of the large scale flow, $\langle w_{LS} \rangle_y /U_{cl,lam}$.}
\end{figure}

\begin{figure}
\centering
\includegraphics[width=\columnwidth]{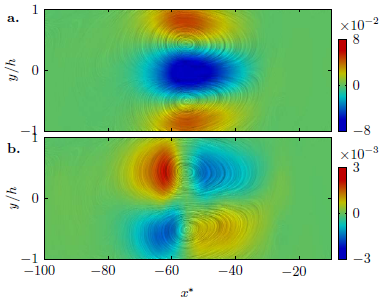}
\caption{\label{fig:LargeScaleMean-xy}Large-scale flow averaged over $-L_z<z<L_z$ for $Re=1500$ textured by LIC technique. \textbf{a.} Streamwise component of the large scale flow, $\langle u_{LS} \rangle_z /U_{cl,lam}$. \textbf{b.} Wall normal component of the large scale flow, $\langle v_{LS} \rangle_z /U_{cl,lam}$.}
\end{figure}

In Lemoult \textit{et al.} \cite{lemoult2013turbulent}, we carried out the decomposition of the fluctuating part of the flow, $\textbf{u}$, into a large-scale flow, $\mathbf{u}_{LS}$, and a small-scale flow, $\mathbf{\tilde{u}}$.
\begin{equation}
\mathbf{u} = \mathbf{u}_{LS} + \mathbf{\tilde{u}}
\end{equation}
In order to compute this large-scale flow, we measured the flow in a wide area and applied spatial filtering. We only measure this large scale flow in the $y=0.5h$ plane. In the present study, we compute the 3D large scale flow field by applying a spatial Gaussian filtering of standard deviation $2h$ in the $x$ and $z$ direction. In order to increase the signal-to-noise ratio, we averaged the velocity field over 10 realisations of the same experiment. We obtain the large-scale flow represented in figures \ref{fig:LargeScaleMean-xz} and \ref{fig:LargeScaleMean-xy}.

Figure \ref{fig:LargeScaleMean-xz} shows the large-scale flow, $u_{LS}$ and $w_{LS}$, in the $x-z$ plane averaged over the entire channel height. A Line Integral Convolution (LIC) technique \cite{laramee2004state} is used to highlight the streamlines of the flow field. The $y$-averaged large scale flow in the $x-z$ plane is formed of a quadrupole centred on the spot. 
The turbulent spot produces a partial blockage of the channel and induces this large-scale flow across its borders. This large-scale flow is reminiscent of that observed around defects and inhomogeneities in Rayleigh-B\'enard convection \cite{croquette1986large} and its strength could be probably estimated from the gradient of the spatial phase variable.

In the $x-y$ plane, the phenomenon is slightly different. Figure \ref{fig:LargeScaleMean-xy} presents the large-scale flow, $u_{LS}$ and $v_{LS}$, in the $x-y$ plane $z$-averaged over the entire channel width. The main difference is $v$ is one order of magnitude smaller than $u$ and $w$.
Again, the partial blockage induced by the turbulent spot creates a large scale flow around it. And the flow is accelerated, compare to the laminar flow, close to the wall. 
This large scale flow, in the $x-y$ plane, forms a dipole centred on the spot.

\begin{figure}
\centering
\includegraphics[width=\columnwidth]{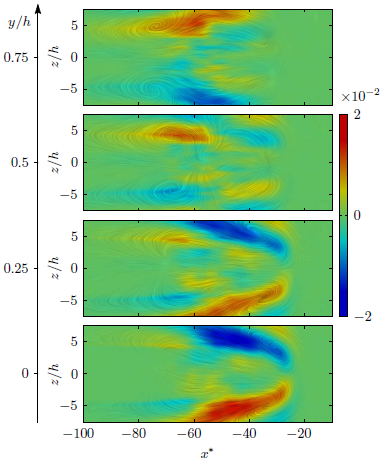}
\caption{\label{fig:LargeScaleMean-y}Large scale wall normal vorticity $\Omega_y$ flow in the $y=0,0.25h,0.5h\mbox{ and } 0.75h$ planes for $Re=1500$ textured by LIC technique using the large scale flow field.}
\end{figure}

In order to better capture the nature of the large-scale flow, we compute the wall normal vorticity associated to the large scale flow $\Omega_y = \partial u_{LS}/\partial z - \partial w_{LS}/\partial x^*$.
Figure \ref{fig:LargeScaleMean-y} presents $\Omega_y$, non-dimensionalized by $U_{cl,lam}/h$ , in four different $x-z$ planes: $y=0,0.25h,0.5h\mbox{ and } 0.75h$, textured with LIC using the large-scale flow.
From this representation it is noticeable that the quadrupole observed in figure \ref{fig:LargeScaleMean-xz} is in fact a 3D structure, mainly composed of the superposition of three dipoles. One is located in $y=0$ and is positioned at the leading edge of the spot whereas the other ones, located near $y= \pm 0.75h$, are located at the trailing edge of the spot.
%
The present study of the large scale flow associated with a turbulent spot confirmed the presence of a quadrupole centred on the spot observed by Lemoult \textit{et al.} \cite{lemoult2013turbulent} and give a description of the variation of this large scale flow along the wall normal coordinate.

\subsection{Low order model}

The TR-SPIV in a cross channel plane gives us the instantaneous 3D flow in and around a turbulent spot. We can then study the dynamics of this turbulent spot in terms of a low-order model. We follow the idea of Barkley \cite{barkley2011simplifying} who models pipe flow as a generic excitable and bistable medium. His two-variable model captures qualitatively the main features of pipe flow transition, including metastability of puffs at low $Re$, splitting of puffs at intermediate $Re$ and slugs at higher $Re$. The model only includes two variables, $(u,q)$ where $u$ is the center line streamwise velocity and $q$ is related to the intensity of the turbulence.

\subsubsection{Continuous model}

We start with a set of two partial differential equations $(u,q)(x,t)$ where $x$ is the streamwise coordinate of the channel which is similar to the continuous model presented by Barkley \cite{barkley2011simplifying} (we will now refer to this model as B11) except we made a substitution in the $u$ variable.

\begin{equation}
\partial_t u + u \partial_x u = \epsilon_1 (1-u) - \epsilon_2 (u - u_{bulk})q - \partial_x u
\label{eq:Model-u}
\end{equation}
\begin{equation}
\partial_t q + u \partial_x q = q \left[ u-1+r-(r+\delta)(q-1)^2 \right] + \partial_{xx} q
\label{eq:Model-q}
\end{equation}

In B11, $u$ is the centerline velocity relative to the mean velocity, we prefer to use the centerline velocity instead. Another change is the advection by the velocity $u$, in the left hand side, instead of advection by a constant velocity in B11. 

\begin{figure}
\centering
\includegraphics[width=\columnwidth]{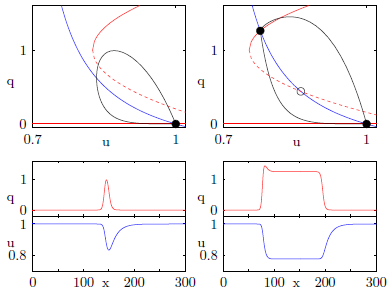}
\caption{Top row: Phase space $(u,q)$ for $r=0.9\ r_c$ (left) and $r=1.2\ r_c$ (right). $u$-nullcline appears in blue and $q$-nullclines appear in red, black line is the result of the model Bottom-row: $u(x)$ and $q(x)$ for the same $r$ than above. \label{fig:NullClines}}
\end{figure}

The core of the model is seen in the $u-q$ phase space represented in figure \ref{fig:NullClines}. We represent the nullclines, \textit{i.e.} curves where all spatial and temporal derivatives are equal to zero. Whatever $r$, there exists a fixed point in $(1,0)$, which corresponds to the laminar parabolic profile. The dynamics of $u$ is quite simple. In the absence of turbulence, $q=0$, $u$ relaxes to $u=1$ at rate $\epsilon_1$, while if $q>0$, $u$ decreases to $u_{mean}=2/3$ at a faster rate dominated by $\epsilon_2$.
There exists two different $q$-nullclines. The $q=0$ curve means that turbulence can not be generated spontaneously from laminar flow but a minimal perturbation is necessary. The second $q$-nullcline is the quadratic curve defined as

\begin{equation}
q = 1 \pm \sqrt{\frac{u-1+r}{r+\delta}}
\end{equation}

The position of its nose, $(1-r,1)$, is controlled by the parameter $r$, which plays the role of a Reynolds number, while it always cut the $q=0$ curve at $u=1+\delta$. The upper branch is attractive, while the lower branch is repelling and sets the nonlinear stability threshold for laminar flow. If laminar flow is perturbed beyond the threshold (which decreases with $r$ like $r^{-1}$), $q$ is nonlinearly amplified and $u$ decreases in response.

While $r < r_c = \epsilon_2/[3(\epsilon_1 + \epsilon_2)]$, there is only one fixed point and the system is excitable. The upstream side of a puff is a trigger front where abrupt laminar to turbulent transition takes place. However, turbulence cannot be maintained locally following the drop in the mean shear. The system relaminarizes on the downstream side whose speed is set by the upstream front. In this regime, turbulent puffs are advected downstream without any change in their shape.
On the other hand, for $r > r_c$, a second fixed point appears. The system becomes bistable and turbulence can be maintained indefinitely by the modified mean shear. The upstream and downstream front move at different speeds and the turbulent region expands. This regime corresponds to the "slug" regime (here slug should be understood as a spreading spot in a 1D PPF).

We compare this model to our experimental data, $U_{cl}$ and $E_{yz}$, where $U_{cl}$ is the streamwise velocity in the centre of the channel and $E_{yz}$, defined as the energy associated with the fluctuating cross flow, is calculated as

\begin{eqnarray}
E_{yz} = \frac{1}{2} \int_{-Lz}^{Lz} \int_{-h}^{h} ( v^2 + w^2 ) dy dz
\end{eqnarray}

\begin{figure}
	\centering
	\includegraphics[width=\columnwidth]{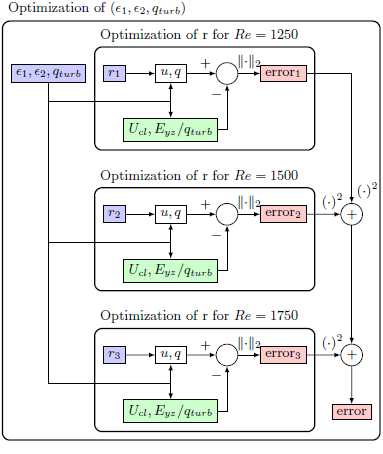}
	\caption{Optimization procedure to identify the five adjustable parameters (blue) by minimization of the error with respect to the experimental data (green).}
	\label{fig:Optimization}
\end{figure}

We extract $U_{cl}$ and $E_{yz}$ from our experimental data for three $Re$ (1250,1500 and 1750). In the model, whereas $u$ is physical and can be compared directly to $U_{cl}$, $q$ is of the order of $O(1)$. In the experimental data, $E_{yz}$ is much smaller. We introduce $q_{turb}$ to  rescale $E_{yz}$ in order to obtain $q = E_{yz}/q_{turb} = O(1)$. In order to identify the parameters which are the most relevant in the channel flow case, we set up an optimization procedure. In addition to $q_{turb}$, there are 5 parameters of the model to find. Two of them, $\epsilon_1$ and $\epsilon_2$, are fixed for all $Re$ and we have to find $r_1$, $r_2$ and $r_3$, the parameter $r$ in each $Re$ cases. We have chosen to set $\delta=0.1$ as proposed by Barkley \cite{barkley2011simplifying}. This process is represented on figure \ref{fig:Optimization} and can be explained as follow. We start with an initial guess for $(\epsilon_1,\epsilon_2,q_{turb})$, then we optimize the value of $r$ for each value of $Re$. In order to achieve this optimization, we use the \texttt{fminbnd} function in Matlab. This algorithm is based on a golden section search and parabolic interpolation and minimizes a function of one parameter on a given interval. The function we minimize is the error between the value of $(u,q)$ predicted by the model and 10 realisations of the experiment. Finally, we use the \texttt{fminsearch} function of Matlab (which uses the Nelder-Mead simplex algorithm as described in Lagarias \textit{et al.} \cite{lagarias1998convergence}) to find the trio $(\epsilon_1,\epsilon_2,q_{turb})$ which minimizes the sum of the square of the individual errors.

\begin{figure}
\centering
\includegraphics[width=\columnwidth]{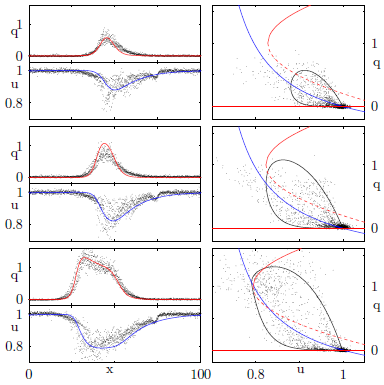}
\caption{\label{fig:Comparison}Top row: $Re=1250$, middle row: $Re=1500$ and bottom row: $Re=1750$. Solid line is the result of the model and dots correspond to experimental data. Left column: $u(x)$ and $q(x)$ for each $Re$, blue lines represent $u$, red lines represent $q$ and dots are 10 experimental realisations. Right column: phase space $(u,q)$, $u$-nullcline appears in blue and $q$-nullcline appear in red. }
\end{figure}

The results of the optimization procedure are presented on figure \ref{fig:Comparison}. For each $Re$ (top row: $Re=1250$, middle row: $Re=1500$ and bottom row: $Re=1750$), we represent 10 experimental realisations (dots, each realisation is a time series of 1500 samples) of $U_{cl}$ and $E_{yz}/q_{turb}$. The solid lines are $u$ and $q$ predicted by the model. We have found that $\epsilon_1=0.1$, $\epsilon_2=0.16$ and $q_{turb}=8\times 10^{-4}$ give the best fit. We also have identified $r_1=0.84$, $r_2=0.87$ and $r_3=1.01$. As expected, $r$ increases with $Re$, even though the relation is not linear. The agreement between the experimental data and the model is quite good and it is important to note that it does not depend strongly on the choice of parameters.

\subsubsection{Additive noise}

While the continuous model captures well the basic properties of the transition process, ``equilibrium'' spots (with constant shape and size) and expanding spots, it is too simplistic to model the abrupt decay of spots or the appearance of bands. We continue to follow the idea of Barkley by adding some noise to the model \cite{barkley2011modeling}. We replace equation (\ref{eq:Model-q}) by the following

\begin{equation}
\begin{split}
\partial_t q + u \partial_x q = q \left[ u-1+r-(r+\delta)(q-1)^2 \right]\\
 + \partial_{xx} q + \sigma q \eta
\label{eq:Model-q-Noise}
\end{split}
\end{equation}
where $\eta(x,t)$ is Gaussian noise. This is exactly the same equation as (\ref{eq:Model-q}) except we add noise proportional to $q$ to the right-hand side of the equation. We also add a new parameter $\sigma$ which controls the strength of the noise. In the following, we set $\sigma=0.08$.

\begin{figure}
\centering
\includegraphics[width=\columnwidth]{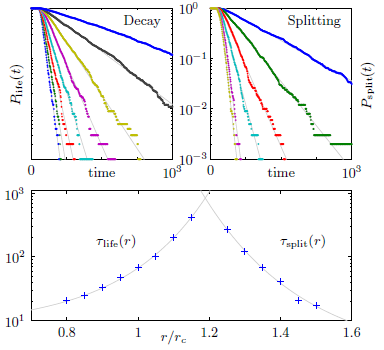}
\caption{\label{fig:DecaySplitting}Top-left: Probability for a puff to survive until a given time, $P_{\rm life}(t)$, for $r=0.8, 0.85, 0.9, 0.95, 1, 1.05 \mbox{ and } 1.1$. Solid lines correspond to the fit of $P_{\rm life}(t)$ by an exponential. Top-right: Probability for a puff to have been splitted at a given time, $P_{\rm split}(t)$, for $r=1.25,1.3,1.35,1.4,1.45 \mbox{ and } 1.5$. Solid lines correspond to the fit of $P_{\rm split}(t)$ by an exponential. Bottom: Mean time life, $\tau_{\rm life}$, and mean splitting time, $\tau_{\rm split}$, calculated from the exponential fit. Solid lines are fit by a super exponential.}
\end{figure}

By adding this noise we allow a spot to relaminarize spontaneously or to split into two distinct spots. In order to perform lifetime statistics, we perform 1000 simulations with the same initial condition and for each simulation we run the simulation as long as the spot survives, \textit{i.e.} $\lVert q \rVert >10^{-2}$. We can then define the probability $P_{\rm life}(t)$ for the spot to survive until $t$. Figure \ref{fig:DecaySplitting}, top left, presents this probability for different $r$. The survival probabilities are exponential, $P_{\rm life}(t) \propto\exp(-t/\tau_{\rm life}(r))$ where $\tau_{\rm life}(r)$ is the mean spot lifetime. It is not possible to compare the mean lifetime founded here since to our knowledge there exists no experimental data in the literature for the lifetime of spots in plane Poiseuille flow. Nevertheless this can be compared to the results found in other wall bounded shear flows \cite{bottin1998discontinuous,bottin1998statistical,avila2011onset,shi2013scale}.

It is also possible to generate splitting time statistics, $P_{\rm split}(t)$, for the model with noise. We perform 1000 simulations with the same initial condition and we run the simulation until a splitting event occurs, \textit{i.e.} two $q$ peaks separate by at least $40h$. We present in figure \ref{fig:DecaySplitting}, top right, this splitting probability as a function of time for different $r$. Similarly to the life time statistic, the spliting probability follows an exponential law, $P_{\rm split}(t) \propto\exp(-t/\tau_{\rm split}(r))$ with $\tau_{\rm split}(r)$ the mean time necessary for a spot to split into two spots.

Figure \ref{fig:DecaySplitting}, bottom, compares the mean lifetime, $\tau_{\rm life}$, and the mean splitting time, $\tau_{\rm split}$, with respect to the parameter $r$. As expected, $\tau_{\rm life}$ increases with $r$ and $\tau_{\rm split}$ decreases with $r$. The intersection point is expected to be the onset of sustained turbulence in the sense that it takes less time to split than to vanish. Moreover, solid lines are fits by a super exponential law. This last point is in total agreement with experimental and numerical studies made in pipe flow \cite{avila2011onset}.
Obviously, the dynamic of PPF is expected to be richer since the spot spreads in both the $x$ and $z$ directions and the splitting will most likely occur in the spanwise direction as seen in visualization experiments \cite{carlson1982,alavyoon1986turbulent}. However we observed, at $Re=4000$, islands of low turbulence inside a turbulent spot suggesting a spreading in the streamwise direction.

\subsubsection{2D extension}

\begin{figure*}
\centering
\includegraphics[width=\textwidth]{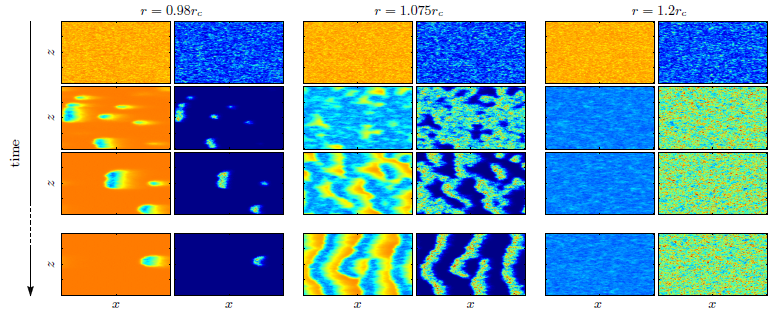}
\caption{\label{fig:SnapshotsBand}Two dimensional simulations of our model with random initial conditions. From top to bottom: $t=0.1,12.5,25 \mbox{ and }100$. For each value of $r$, left column represents $u$ from 0.7 (blue) to 1.1 (red) and right column represents $q$ from 0 (blue) to 2 (red). The size of the domain is $(200h\times 100h)$.}
\end{figure*}

This one-dimensional model is obviously too simple to capture the richness of the transition in plane Poiseuille flow but it allows us to identify relevant parameters and we can now extend this model to two dimensions. We add the spanwise coordinate $z$ in our model, and we simulate $u(x,z,t)$ and $q(x,z,t)$. The equation for $u$ remains unchanged but we add some diffusion in the spanwise direction for $q$. The new system is composed of equation (\ref{eq:Model-u}) and (\ref{eq:Model-q-2D}).

\begin{equation}
\begin{split}
\partial_t q + u \partial_x q = q \left[ u-1+r-(r+\delta)(q-1)^2 \right]\\
 + D_{\parallel}\partial_{xx} q + D_{\perp}\partial_{zz} q + \sigma q \eta
\label{eq:Model-q-2D}
\end{split}
\end{equation}
where $D_{\parallel}$ and $D_{\perp}$ are the coefficients of diffusion in the streamwise and spanwise directions and $\eta(x,z,t)$ is Gaussian noise.

Unlike in plane Couette flow, due to the advection of the turbulent spot by the mean flow, it is difficult to obtain experimental statistics on the mean lifetime or splitting time of spots in plane Poiseuille flow. There are only few studies which mention the splitting of turbulent spots in plane Poiseuille flow \cite{carlson1982,alavyoon1986turbulent}. These studies do not present any systematic statistics on the spot, but they only report the observation of "equilibrium" spots or split spots. One important characteristic of spots is the V-shape that is easily observable after few times. From a numerical point of view, DNS of large aspect ratio Poiseuille flow are costly in term of computer time. In consequence most of numerical studies are only concerned with an isolated spot \cite{henningson1987numerical,henningson1991turbulent}. However, more recently,the question of the existence of turbulent bands in plane Poiseuille flow has received attention. Aida \textit{et al.} \cite{tsukahara2011} performed a DNS of transitional plane Poiseuille flow in a very large domain $(730h\times 2h\times365h)$ and observed the development of a turbulent spot in the forms of two arms growing in the $x-z$ plane with an angle of approximately $\pm45^\circ$ and then they observed the appearance of turbulent bands. Tuckerman \cite{tuckerman2013ETC14} used the tilted domain technique \cite{barkley2005computational} to observe the formation of turbulent bands.

Due to the lack of statistics on turbulent spots in PPF, we will just check if our model is able to mimic some general features of the transition to turbulence: localized spots, turbulent bands and featureless turbulence.

We start simulations in a $(200h\times 100h)$ wide domain with periodic boundary conditions in both spanwise and streamwise direction. Initially, $u$ is set to 1 uniformly and $q$ is set to a random initial condition in the entire domain. Snapshots of those simulations are presented on figure \ref{fig:SnapshotsBand} for different values of $r$. For small $r$, only a few localized spots survive. After they had been generated, these spots follow the mean lifetime statistics independently and disappear suddenly at different times. Finally, at $t=100$, only one spot survives. At high $r$, we observe that the entire domain becomes turbulent and that the mean of $u$ falls to 0.8. This regime is comparable to featureless turbulence. The most interesting regime occurs at intermediate $r$. In this regime, we observed the formation of alternated turbulent and laminar bands. These bands form an angle with respect to the streamwise direction of approximatively $\Theta \approx 50^\circ$ in a relative good agreement with Duguet \textit{et al.} \cite{duguet2010formation} in plane Couette flow. However if the simulation is carried out for larger time, the bands tend to form an angle of $\Theta=90^\circ$. The space between bands is largely governed by $D_{\parallel}$. This is due to the diffusion term in the spanwise direction which tends to make $q$ uniform in this direction. To avoid this phenomenon one idea is to use the large scale flow induced around a localized turbulent spot (see \S\ref{par:LSflow}) as a driving mechanism of the spreading of spot. This idea has been suggested by Duguet \textit{et al.} \cite{duguet2013oblique} in plane Couette flow. In future work we will add to this model the spanwise velocity, $w$, as a third variable.

\section{Conclusion}

We have performed new precise measurements of the three components of the flow in and around a turbulent spot in transitional channel flow. We have been able to observe for the first time travelling-wave-like structures close to the trailing edge of a spot. This observation supports the idea of dynamical systems theory that these exact coherent structures may indeed capture the nature of fluid turbulence. We also report a precise description of the large scale flow associated with the turbulent spot. We confirmed that this flow consists of a quadrupole centred on the spot and give a description of its variation along the wall normal coordinate.

Finally, starting from the continuous model of pipe flow proposed by Barkley \cite{barkley2011simplifying}, we have built a set of two coupled non-linear equations for two variables, the center line velocity and the turbulence intensity, which captures the main features of transitional plane Poiseuille flow. We have been able to mimic the appearance of turbulent bands in plane Poiseuille flow.

\begin{acknowledgement}

G.L. would like to thank the DGA for its support. K.G would like to thank the ESPCI for its support through the Joliot Curie Chair. Dwight Barkley is gratefully acknowledged for helpful discussions and sharing his experience on his previous work. Laurette Tuckerman is acknowledged for valuable discussions. Finally, EPJE is acknowledged for giving us the opportunity to write this article in a topical issue dedicated to Paul Manneville.
\end{acknowledgement}

\bibliographystyle{unsrt}
\bibliography{biblio}

\end{document}